\documentclass[10pt,journal,compsoc]{IEEEtran}
\ifCLASSOPTIONcompsoc
  \usepackage[nocompress]{cite}
\else
  \usepackage{cite}
\fi

\usepackage{nameref,hyperref}
\usepackage{changepage}
\usepackage{color}
\usepackage[flushleft]{threeparttable}

\usepackage{graphicx}

\usepackage{wrapfig}
\usepackage[pscoord]{eso-pic}
\usepackage[fulladjust]{marginnote}
\reversemarginpar
\usepackage{caption}
\usepackage{subcaption}
\usepackage{listings}
\usepackage{xcolor}

\definecolor{codegreen}{rgb}{0,0.6,0}
\definecolor{codegray}{rgb}{0.5,0.5,0.5}
\definecolor{codepurple}{rgb}{0.58,0,0.82}
\definecolor{backcolour}{rgb}{0.95,0.95,0.92}

\lstdefinestyle{mystyle}{
    backgroundcolor=\color{backcolour},   
    commentstyle=\color{codegreen},
    keywordstyle=\color{magenta},
    numberstyle=\tiny\color{codegray},
    stringstyle=\color{codepurple},
    basicstyle=\ttfamily\footnotesize,
    breakatwhitespace=false,         
    breaklines=true,                 
    captionpos=b,                    
    keepspaces=true,                 
    numbers=left,                    
    numbersep=5pt,                  
    showspaces=false,                
    showstringspaces=false,
    showtabs=false,                  
    tabsize=2
}

\usepackage{tikz}
\usetikzlibrary{decorations.pathreplacing,calc}

\usepackage{amsmath}
\usepackage{calc}
\newcommand{\explain}[2]{\underbrace{#1}_{\parbox{\widthof{\ensuremath{#1}}}{\footnotesize\raggedright #2}}}
\renewcommand*\descriptionlabel[1]{\hspace\leftmargin$#1$}

\newcommand{\Data}{\texttt{Data} }
\newcommand{\Interest}{\texttt{Interest} }
\newcommand{\Fig}{Fig. }

\usepackage{algorithm}
\usepackage{algpseudocode}
\usepackage{float}

\usepackage{rotating}
\usepackage{amsthm}
\usepackage[safe]{tipa}
\newtheorem*{remark}{Remark}

\begin{document}

\title{R2: A Distributed Remote Function Execution Mechanism With Built-in Metadata}

\author{Jianpeng~Qi, and~Rui~Wang,~\IEEEmembership{Member,~IEEE,}
        \IEEEcompsocitemizethanks{\IEEEcompsocthanksitem This paper was
        published in IEEE. DOI: 10.1109/TNET.2022.3198467 \protect\\
}
}

%
%

\markboth{Journal of IEEE Transactions on Networking,~Vol.~\#, No.~\#, August~2021}%
{Jianpeng \MakeLowercase{\textit{et al.}}: R2}
%



\IEEEtitleabstractindextext{%
\begin{abstract}
  Named data networking (NDN) constructs a network by names, providing a
  flexible and decentralized way to manage resources within the edge computing
  continuum. This paper aims to solve the question, ``Given a function with its
  parameters and metadata, how to select the executor in a distributed manner
  and obtain the result in NDN?'' To answer it, we design R2 that involves the
  following stages. First, we design a name structure including data, function
  names, and other function parameters. Second, we develop a 2-phase mechanism,
  where in the first phase, the function request from a client-first reaches the
  data source and retrieves the metadata. Then the best node is selected while
  the metadata responds to the client. In the second phase, the chosen node
  directly retrieves the data, executes the function, and provides the result to
  the client. Furthermore, we propose a stop condition to intelligently reduce
  the processing time of the first phase and provide a simple proof and range
  analysis. Simulations confirm that R2 outperforms the current solutions in
  terms of resource allocation, especially when the data volume and the function
  complexity are high. In the experiments, when the data size is 100 KiB and the
  function complexity is $\mathcal{O}(n^2)$, the speedup ratio is 4.61. To
  further evaluate R2, we also implement a general intermediate data processing
  logic named ``Bolt'' implemented on an app-level in ndnSIM. We believe that R2
  shall help the researchers and developers to verify their ideas smoothly.
\end{abstract}

}

\IEEEoverridecommandlockouts
\IEEEpubid{\makebox[\columnwidth]{1558-2566 \copyright2022 IEEE. Personal use is permitted, but republication/redistribution requires IEEE permission.\hfill}
\hspace{\columnsep}\makebox[\columnwidth]{ }}
\maketitle
\IEEEpubidadjcol

\IEEEdisplaynontitleabstractindextext

%
\IEEEpeerreviewmaketitle

\IEEEraisesectionheading{\section{Introduction}\label{sec:introduction}}

\begin{figure}[!th]
  \centering
  \includegraphics[width=70mm]{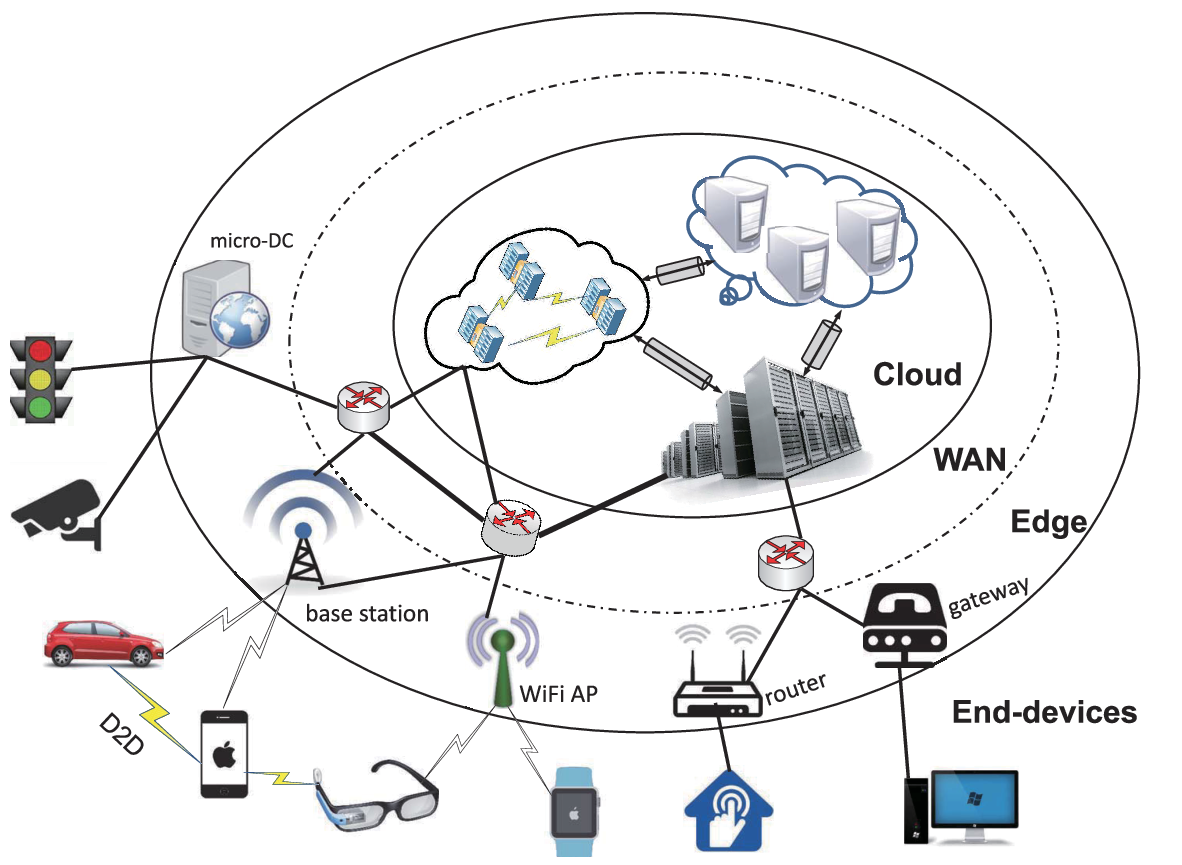}
  \caption{\textit{Illustration of edge computing\cite{Zhou_2019}. }(Copyright © 2019, IEEE)}
  \label{fig:illustration-of-edge-computing}
\end{figure}

\IEEEPARstart{I}{DC estimates}  that by 2025 41.6 billion devices will be
interconnected, and data volume will reach 79.4 zettabytes (ZB)
\cite{Framingham}. Current cloud computing architectures do not afford such an
overwhelming amount of devices and data due to high latency, limited bandwidth,
high carbon footprint, and poor security \cite{Shi_2016}. Many traditional
services that are processed in the cloud but generated remotely
\cite{NguyenMehtaKleinEtAl2019} are thus maintained at a high cost, e.g., the
communication and computation-intensive tasks, such as image recognition and
target tracking. Thus, edge computing, an accelerator of cloud computing,
affords better computing resources for users and thus gains widespread
attention. Current edge computing features, i.e., ultra-low latency,
geographical distribution, unlimited bandwidth, high privacy, and security,
offset the lack of cloud computing and employ it used in many realms
\cite{MouraHutchison2020}. Additionally to the edge resources, as illustrated in
\Fig \ref{fig:illustration-of-edge-computing}, \textit{``edge computing includes
employing networked resources closer to the data sources/sinks. Resources can be
at the edge, in the cloud, and everywhere in between a continuum''}
\cite{Shi_2016, Mor2019, Zhou_2019}. Hence, edge computing affords task
processing and data analysis everywhere.

Edge resources, specifically the continuum, are usually employed to execute
tasks in a distributed manner to provide a time-limited service. For example,
developers may partition the communication and computation-intensive deep
learning models, consecutively deploy them on the continuum, and then execute
them to obtain the results with a short delay \cite{edgeai}. Compared with
purely cloud processing, continuum processing has several benefits: low energy
cost and low latency. However, the latter advantages assume that the
nodes/networks state and the data sources details are known, even for a trivial
counting job, such as task scheduling, resource discovering, and data
retrieving. The most popular way of knowing and managing these states and
details is utilizing a central metadata server that involves abstract data about
essential attributes such as location, size, and format. This strategy is used
in many applications like the ``Metastore'' or ``NameNode'' of Apache Hadoop. By
managing the metadata through the central node, the developers or users can
easily leverage their resources.

Nevertheless, in an edge computing scenario, metadata management is non-trivial,
as developing such a central server is costly. This is because enormous small
data pieces, chunks, and files, are scattered in geo-graphical distribution, and
gathering and managing them in a centralized manner becomes infeasible, as the
nodes within the continuum are usually volatile, highly changeable, and even
unreliable \cite{9409738,10.1145/3464428}. Electing a consensus node with high
availability in such a dynamic environment imposes high scheduling and
maintenance costs \cite{IslamKumarHu2021}, prohibiting edge computing from being
affordable. Furthermore, cost estimation, monitoring the remote nodes’ status,
authentication, and service optimization is challenging. Therefore, utilizing
the edge resources in a decentralized and native way instead of a consensus node
becomes critical.

Some typical applications like CDN and DNS accelerate the network’s data access
speed. Specifically, CDN caches offer little in terms of general computational
capabilities \cite{HarcholMushtaqFangEtAl2020}, while DNS 
\cite{HsuChoncholasBhardwajEtAl2020} introduces additional inevitable name
resolution delays \cite{10.1145/3464428}. Nevertheless, the current IP
architecture suitable for point-to-point communication is limited in
distributed networks \cite{ZhangAfanasyevBurkeEtAl2014}.

Fortunately, Information-Centric Networking (ICN) and, in particular, its
prominent Named Data Networking (NDN) instantiation
\cite{ZhangAfanasyevBurkeEtAl2014} that constructs the network based on the data
name rather than IP provides a realistic solution. Specifically, the work based
on NDN uses the function/service name as its routing rules to find the proper
executor in a fully distributed manner. This strategy affords to provide quickly
and efficiently several scalable and robust services, e.g., serverless
computing, function as a service, and in-network computing
\cite{ScherbFaludiTschudin2017, Mastorakis_2020, KrolHabakOranEtAl2018}.
However, ICN/NDN technology is still at an early stage and requires more
investigation \cite{AmadeoRuggeriCampoloEtAl2019}. 

Many works that integrate the function or service name and directly transmit the
raw data to the executor may cause a blocking problem. Thus, metadata should
solve the latter issue before accelerating the edge computing speed. Besides the
proper executor selection, retrieving metadata as the first step presents
several side benefits, such as checking the node’s health or working status in
the data forwarding path, affording cost estimation by relatively small metadata
to avoid network congestion, and warming up the dependent environment of the
function/service.

Nevertheless, the following question arises, ``is it feasible to perform
in-network processing in NDN according to the metadata?''. To answer this
question, this paper investigates leveraging resources within the continuum in a
distributed fashion to satisfy the function/service requirement in NDN. The
proposed design presents the following features: (1) Data and metadata are
distributively stored, with the preferred location being the data producer. (2)
The user publishes an Interest containing a data name and a function with
parameters and only expects the function result (output of the function). (3)
The function can be executed along the continuum from the user to the data
generator. (4) Most importantly, using metadata to select the best function
execution node to minimize the end-to-end delay.

Based on these features, we propose a 2-phase distributed remote function
execution mechanism utilizing metadata. During the first phase, we select the
best executor in the forwarding path according to the data abstract, node
status, and network condition to minimize the total end-to-end delay. Then, in
the second phase, we use this executor as a ``transit station'' to receive the
large raw dataset, analyze it, and send the short analyzed result to the
requester. We verify this idea on the NDN project
\cite{ZhangAfanasyevBurkeEtAl2014}, while most codes are implemented on an
app-level to make R2 computing scalable. Since the user only cares about the
result, we name our method ``Request Result'' (``R2''). The major contributions of
this work are as follows:
\begin{enumerate}
\item	 We propose R2 to prove the feasibility of distributively running the
       function based on metadata. R2 assumes the metadata is stored together
       with its data and utilizes a 2-phase mechanism to consecutively complete
       the metadata extraction, cost estimation, function executor selection,
       data extraction, function execution, and result response.
\item  To find the best node for function execution, we formulate a
      distributive cost estimation process based on metadata combined with a stop
      condition. This strategy avoids scanning all nodes along the forwarding path to
      reduce the result retrieving time automatically. The stop condition can also be
      used in other uncertain edge computing scenarios to distributedly select the
      best node.
\item  We implement, evaluate and analyze the performance of R2 on ndnSIM
       \cite{mastorakis2017ndnsim}, involving numerous experiments conducted on
       a real-world network topology dataset. Additionally, to make R2 flexible
       and portable, we developed an application-level intermediate data
       processing plugin named ``Bolt'', which can be installed on a
       computing-capable node to make R2 scalable. R2 is open source and can be
       found on GitHub \cite{r2sourcecode}.
\end{enumerate}

The remainder of this paper is as follows. Section \ref{sec:related-work}
presents the related work on computing within the continuum. Then Section
\ref{sec:r2} introduces some preliminaries of NDN and the details of the R2,
including name format, forwarding pipeline, cost optimization, applied scenes,
and proofs. Section \ref{sec:evaluation-and-results} discusses the experiments,
numerical results, and the ``Bolt'' implementation, while Section
\ref{sec:discussion} discusses some security issues. Finally, Section
\ref{sec:conclusion} concludes this paper.

\section{Related Work}\label{sec:related-work} %
\begin{table}[th]
  \centering
  \renewcommand{\arraystretch}{1.3}
  \begin{adjustwidth}{0in}{0in} 
  \centering
  \caption{Comparison with the notable related works}
  \label{tbl:comparison-of-recent-works}
  \begin{center}
    \begin{tabular}{p{0.1\columnwidth} p{0.15\columnwidth} p{0.15\columnwidth}
    p{0.2\columnwidth} p{0.15\columnwidth}}
      \hline
      Work$^1$ & Forwarding strategy & Executing level &
      Distributed executor selection & Executor selection space$^2$\\ [0.5ex]
      \hline\hline
      NFaaS \cite{KrolPsaras2017}  & Custom & NFD & None (Service duplicating) &
      A $\bigcap$ E\\ 
      \hline
      RICE \cite{KrolHabakOranEtAl2018}  & Builtin  & App & None & Producer \\
      \hline
      CFN \cite{KrolMastorakisOranEtAl2019} & Builtin  & App & Moderate (Task
      scheduler) & A $\bigcap$ E \\
      \hline
      NDNe \cite{Amadeo_2016} & Builtin  & NFD & Distributed (Replies first) & A
      $\bigcap$ E \\
      \hline
      IoT-NCN \cite{AmadeoRuggeriCampoloEtAl2019}  & Custom  & NFD & Distributed
      (Piggybacking) & P $\bigcap$ E \\
      \hline
      ICedge \cite{Mastorakis_2020}  & Custom & App & Distributed (Monitoring
      metrics) & A $\bigcap$ E  \\
      \hline
      \cite{10.1145/3464428}  & Custom  & App & Distributed (Monitoring metrics)
      &  A $\bigcap$ E  \\
      \hline
      R2  & Custom & App & Distributed & P $\bigcap $ E\\
      \hline
   \end{tabular}
   \begin{tablenotes}
    \footnotesize 
    \item 1. \textbf{A}: Nodes matched the given name (prefix); \textbf{E}:
    Nodes that can execute the function; \textbf{P}: Nodes on the forwarding
    path. 
    \item 2. CFN, NDNe, ICedge, and \cite{10.1145/3464428} send the function
    name to the executor at first.
  \end{tablenotes}
   \end{center}
  \end{adjustwidth}
\end{table}

Executing functions within the continuum is a well studied topic in many realms,
such as in-network caching \cite{Ullah_2020}, in-network computing
\cite{Sapio_2017}, software-defined networking (SDN) \cite{SonHeBuyya2019}, and
network function virtualization (NFV). ICN is a straightforward way of
forwarding the users’ interest without a centralized coordinator and considering
name resolving, and thus in this paper, we focus on the related work utilizing
ICN. Table \ref{tbl:comparison-of-recent-works} compares the notable related
works regarding the implementation level of the forwarding strategy, the
function executing logic, and the selection of the function/service executor. It
should be noted that, in addition to the comparable items in Table
\ref{tbl:comparison-of-recent-works}, our work provides a simple distributed way
to check the existence of the requested data instantly and minimize the
end-to-end delay by using metadata.

Combining ICN and edge computing has inspired many in-network processing works.
According to their design goal, these works can be generalized into either
service routing or best node selection. Routing function/service according to
the name, such as Named Function as a Service (NFaaS) \cite{KrolPsaras2017} and
NDNe \cite{Amadeo_2016}, can support many edge-native services. Thus, NFaaS and
NDNe are typical serverless computing methods allowing cloud computing to jump
into edge computing, where instead of a centralized metastore, the function name
is employed to fabricate a decentralized service network. This strategy affords
a more scalable and flexible distributed network using the function name as its
routing policy. Michał et al. propose a 4-way handshake remote method invocation
(RICE) \cite{KrolHabakOranEtAl2018} in NDN. In RICE, the consumer first sends a
function Interest (I1) carrying a handshake identifier to the producer that runs
the function and creates a reverse path from the producer to the consumer.
Second, when the producer receives I1, it creates an Interest (I2) containing
the received identifier, following the previously established reverse path
towards the client. Third, the client responds and sends the parameters (D2)
after receiving I2. At last, the producer executes the function with its input
parameters D2 and responds to the result (D1) concerning I1. By integrating a
4-way handshake design, RICE solves several issues, including timer and privacy
concerns. However, it is not handled how to select the proper node to run the
function. 

Memory and computing capacity decrease as we descend levels within the continuum
and move closer to the client \cite{MortazaviSaleheGomesEtAl2017a}. Thus,
selecting the best node is vital within the continuum, especially for critical
applications. Michał et al. further propose the Compute First Networking (CFN)
scheme that relies on RICE to solve the node selection issue
\cite{KrolMastorakisOranEtAl2019}. Forwarding in the store-and-forward network
like NDN affords the same function to be cached everywhere. Thus, ``compute
reuse'' can also be utilized to avoid re-computing the same tasks in a
multi-user scenario \cite{Mastorakis_2020}. Marica et al. propose an IoT-Named
Computation Networking (IoT-NCN) framework \cite{AmadeoRuggeriCampoloEtAl2019}.
This method estimates the service cost by \textit{piggybacking} a
SERVICEEXECCOST field in the Interest and updates the SERVICEEXECCOST value in a
distributed manner. After the Interest reaches the last edge node, an
\textit{executor acknowledgment} is sent back to finish the best executor
election.

However, in a data-centric network, we argue that the data name should also be a
component of the Interest name. Most of the above-mentioned service-oriented
methods, except for IoT-NCN, assume the routing strategies rely on the function
name rather than the data name, which is not capable of generalizing the case
where data is stored on the other edge node. At least additional Interest needs
to be sent. IoT-NCN \cite{AmadeoRuggeriCampoloEtAl2019} adds the data name in
the Interest in front of the function name and routes it to the last edge node
on the IoT domain to create data-oriented distributed services. However, its
cost estimation does not integrate with metadata, i.e., it is ineffectual to the
actual data size and type, especially for other data analytic applications.

Unlike current works, R2 overcomes the challenges mentioned above by utilizing
2-phase operations with metadata. R2 uses the data name as its forwarding strategy
to satisfy the data-oriented applications and metadata to enhance the best node
selection accuracy.

\section{R2}\label{sec:r2}

\begin{figure}[!th]
\centering
\includegraphics[width=75mm]{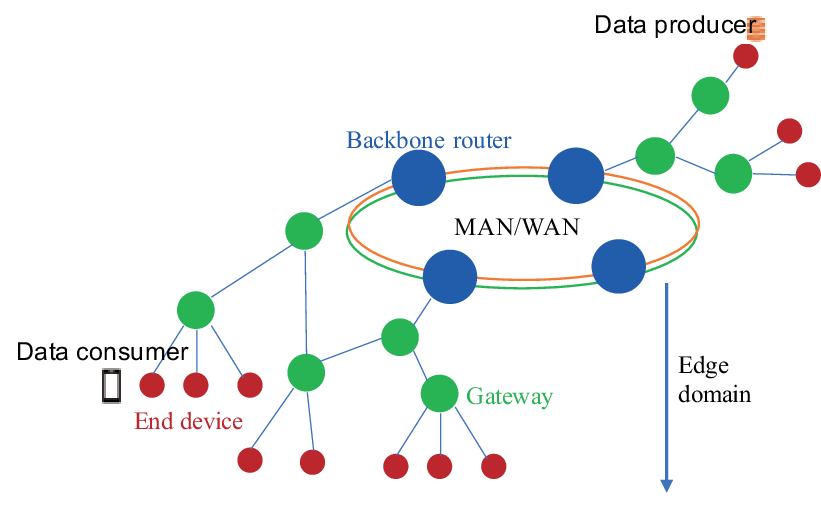}
\caption{Reference network topology.}
\label{fig:reference-network-topology}
\end{figure}
This section first provides the reference topology illustrated in \Fig
\ref{fig:reference-network-topology}. This topology is tangible, especially in a
cross-edge analytic area \cite{Jin_2021}, where the backbone router can
represent an edge site covering a city or a town, and the gateway can be an edge
server covering a community or a home. The available computing (bandwidth)
capacities of these nodes (links) decrease (increase) as we descend levels and
move closer to the end device \cite{MortazaviSaleheGomesEtAl2017a}. We assume
the data with their metadata are commonly stored in the end device. It is worth
noting that due to the ISP barrier, the number of hops between the two mutually
reachable end devices in different neighboring cities can be significant
\cite{XiangWangChenEtAl2019}. Thus, it is mandatory in such an unstable
environment to find the best node within the continuum to perform the user’s
function.

Section \ref{sec:ndn-preliminaries} provides some NDN preliminaries and a
straightforward example to clarify our ideas. Then, we give an elucidation of
the R2 process (Section \ref{sec:R2-process}), including name structure (Section
\ref{sec:name-structure}), cost model and the best node selection (Section
\ref{sec:cost-model-and-select-executor}), the stop condition of reducing delay
(Section \ref{sec:stop-condition}), proofs, and some analysis (Section
\ref{sec:applied-range-analysis}).

\subsection{NDN preliminaries}\label{sec:ndn-preliminaries}
As a concrete architecture of ICN, the NDN project
\cite{ZhangAfanasyevBurkeEtAl2014} aims to solve the communication network
problem. In a communication network (e.g., IP network), packets are named-only
endpoints, and thus it is hard to apply them to the edge computing area where
numerous devices exist. Instead of IP, NDN uses the Name concept to fabricate
the network among different nodes. This architecture contains two basic  
network data packets, Interest and Data. An Interest packet comprises the Name
of the requested data and other options defined by the requester. Data consists
of the data Name, data contents, signature, and other user-defined information.
When a user (also called consumer in NDN) retrieves the data, he first sends an
Interest packet into the network. Then the NDN forwarder (NDN Forwarding Daemon,
NFD) enters the forwarding pipeline to redirect Interest toward the data
producer according to the Interest name. Finally, the producer responds to the
requested data back to the consumer. The consumers drive communication in NDN,
i.e., a receiver-driven communication mode. Other techniques, e.g.,
session support \cite{GasparyanMarandiSchillerEtAl2019, 10.1145/3464428}, and
Long-Lived Interest, can also ease data pushing from the producer.

The NDN design assumes hierarchically structured names. For example, it
retrieves the camera stream or the file \texttt{remote-monitor-data} stored in
Alice’s home through NDN to give a further analysis like confirming her baby is
safe. The Name of Interest might be \texttt{/alice's-home/remote-monitor-data},
where \texttt{/} delineates name components in text representations. Each node in
NDN first checks its Content Store (CS), then Pending Interest Table (PIT) and
Forwarding Information Base (FIB), and finally redirects this Interest by the
Forwarding Strategy to the data node located at Alice's home.\footnote{CS is a
container to store Data packets, PIT is a list to store the unsatisfied Interest
packets, FIB is a routing table which maps Name components to next hops, and
Forwarding Strategy is a series of policies and rules about forwarding \Interest
and \Data packets.} This process is commonly adopted in a longest-prefix
matching method.

\noindent\textbf{An example. }{Recently, many homes where families have babies
or are feeding pets have installed several types of systems to monitor and
control their security devices remotely (using a smartphone and an app). These
systems usually contain machine/deep learning models to provide intelligent
computation and communication-intensive services such as fall detection, fire
warning, and home security checking. Some systems may need time to identify the
most critical event, e.g., the alarm system. A common feature of these jobs is
the a priori known dataset schema, as different schemas typically have different
processing logic. In Alice’s example, the camera types and video format can be
different, such as 60 fps at 12K, 110 fps at 8K, or 220 fps at 4K. Hence,
different function adaptors are needed during runtime. In addition to data
schema, data size is another vital indicator of the service delay (in Alice’s
example, the data volume is enormous). Transmitting these massive HD video
streams identified by \texttt{/alice's-home/remote-monitor-data} on the entire
network could be challenging for the backbone and Alice’s internet expenditure.

Indeed, Alice cares about her baby’s position, and thus, we need to find a
trusted node\footnote{%
A series of trusted nodes can be selected by many techniques such as
blockchain consensus, secure multi-party computation, or manually deploying
authentication services in advance.} %
to identify or analyze the baby’s status, i.e., get the baby’s \texttt{position}
or extract one picture having the baby by running a \texttt{detect} function in
the HD video with a short delay. Finding the node requires the metadata of the
\texttt{remote-monitor-data}, such as the data size, format, and resolution. This
information assists in making a more accurate and fast choice on minimizing the
analysis delay before transmitting the large file. In this case, metadata
content can be in a JSON format %
\texttt{\{resolution:12k, filetype:AVI, size:12MiB\}}, with the attribute
\texttt{size} being the most important. Furthermore, some devices have
flashcards that are incapable of storing the entire dataset, and thus additional
hard disc I/O operations are involved. Nevertheless, metadata information can
help us avoid these devices.}

Next, according to Alice's example, we give the name structure in R2.

\subsection{R2 Interest name structure}\label{sec:name-structure}

\begin{figure}[!th]
  \centering
  \includegraphics[width=85mm]{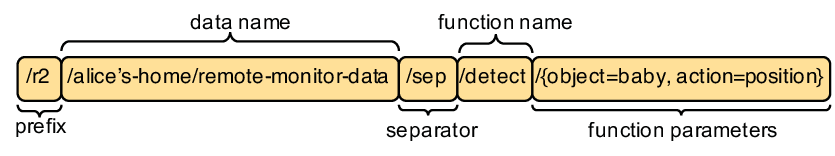}
  \caption{Interest name structure.}
  \label{fig:interest-packet-structure}
\end{figure}

Following Alice's example as mentioned before, \Fig
\ref{fig:interest-packet-structure} depicts the entire \Interest
\texttt{Name} in R2. 

Name in \Fig \ref{fig:interest-packet-structure} means finding the data that
named by \texttt{/alice's-home/remote-monitor-data} and obtaining the baby’s
\texttt{position} by running the \texttt{detect} function. It starts with a prefix
\texttt{/r2}  identification used to avoid mixing with the traditional NDN
Interest. Then, we place the \texttt{/alice's-home/remote-monitor-data} components
behind \texttt{/r2} to let NFD daemons find the camera in Alice’s home. We put the
function component \texttt{/detect} behind data-name and separate them with
\texttt{/sep}. Component \texttt{\{object = baby, action = position\}} involves
the \texttt{ApplicationParameters} of NDN Interest

\begin{remark}
Name matching strategies of NFD usually recognize the Interest by the data name
in a ``Longest Prefix Match'' fashion. Here the data name is
\texttt{/alice's-home/ remote-monitor-data} not \texttt{/r2}. This difference is
solved by the proposed  ``Bolt''and a custom Strategy. Other techniques, such as
``Forwarding Hint'', can be also used. In general, ``Bolt'' is designed on an
application level can perform cost estimation, function execution, and states
reservation. If users or developers want to run R2 on a computing-capable node
such as base stations, gateways, micro data centers and relatively powerful
routers, they can simply install the ``Bolt'' app on the node instead of injecting
junk codes or some unexpected behavior into NFD.
\end{remark}

\subsection{R2 process: a 2-phase precise remote method invocation mechanism}
\label{sec:R2-process}
Here, we re-emphasize the kernel question: \textbf{Given a function with its
parameters and the involved metadata, how to select the best node (i.e., the
executor) in a distributed manner and obtain the final output result?} In other
words, how to select an executor in the forwarding path to \texttt{detect} Alice's
baby's status.

Before answering this question, three rudimentary points need to be addressed.
(1) The requested data should exist, i.e.,
\texttt{/alice's-home/remote-monitor-data} should exist when Alice wishes to
access it, or analysis even by hand is impossible. This point ensures the user a
more reliable service.
(2) The function cost (for the remainder of the paper, ``delay'' shall be a
substitute for ``cost'') can be easily estimated. The executor can be quickly
chosen based on the function \texttt{detect} and the metadata
\texttt{\{resolution:12k, filetype:AVI, size:12MiB\}}, the executor can be
quickly chosen. We use metadata Interest that requests to solve these two
points. 
(3) The executor can be independently, automatically, and distributively
selected based on the first and the second point. Some traditional solutions
armed with consensus nodes intend to deploy related components or services in
the fixed edge nodes (e.g., the edge cloud, which can be regarded as wireless
base stations) placed at the level of the network backbone may not be an optimum
choice. These solutions fulfill the user’s tasks by gathering changeable
information but may be trapped in the consensus problem, as the changeable
information is not easy to synchronize. The third point refers to selecting the
executor based on metadata without requiring additional coordination among nodes
or a coordinator and that the metadata is only the information carrier.

Thus, we design the R2 protocol comprising a 2-phase architecture. The first
phase neglects those 3 points by selecting the executor through cost estimation
based on the function and the retrieved metadata. Developers or maintainers can
also start to warm the runtime environment in this phase. The second phase
involves executing the function on the executor based on the retrieved Data and
then replying the result to the client. Thus, two Interest types are designed,
the metadata-Interest and result-Interest, respectively. Technically, compared
to metadata-Interest, the result-Interest carries a
\texttt{MinCostMarker}\footnote{\texttt{MinCostMarker} tag can be found on
https://git.io/J3ysv} used to identify the executor. 

The result-Interest might also be sent from a node within the continuum, except
for Alice, which we denote as $b$ (bound) for short. An optimized version of R2
in a step-by-step approach is presented in Section
\ref{sec:step-by-step-approach}, illustrated in  \Fig
\ref{fig:automatically-retrieve-result} (right part), introducing a 2-phase
process with its time slices. It should be noted that this version can
automatically select the executor with minimal cost by jointly considering the
network condition, the data source, and the computation capacity. Next, we
introduce in detail the R2 process, including the objective function of
minimizing the end-to-end delay, algorithms, and proofs. Then we give a
fundamental applied analysis of R2.

\subsubsection{Modeling the cost and selecting the executor}\label{sec:cost-model-and-select-executor}%

\begin{figure}[H]
  \centering
  \includegraphics[width=.45\textwidth]{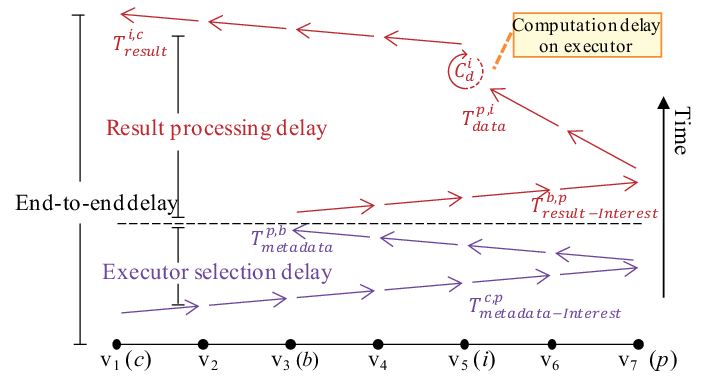}
  \caption{R2 end-to-end delay model.}
\label{fig:2-phase-delay-model}
\end{figure}
\noindent\textbf{Objective function.} Let $V=\{v_1,...,v_N\}$ denote a sequence
of edge nodes along the path from the client $c$ to the producer $p$, $v_i$ is
the chosen $i$th node to execute the function, $C_d^i$ be the computation delay
on $v_i$, and $T^{i,j}_{type}$ be the delay of transmitting the $type$-packet
from $v_i$ onto $v_j$. \Fig \ref{fig:2-phase-delay-model} depicts the delay
model of R2 and includes the executor selection delays (or the cost
estimation, the first phase) and the result processing (the second phase). The
first phase mainly selects the best executor $v_i$ according to the metadata
retrieved by sending an metadata-Interest. The second phase retrieves the raw
data from $p$ by sending an result-Interest on a node $b$ \footnote{We
temporarily use $c$ and $b$ interchangeably. Section \ref{sec:stop-condition}
gives the specific usage of $b$.}, performs the function on the node $v_i$, and
returns the result. Then the end-to-end delay $D$ may be written as:
\begin{align}\label{eq:total-delay}
  D = & \explain{T^{c,p}_{metadata-interest} + T^{p,b}_{metadata}}{cost estimation delay} + \nonumber\\
      & T^{b,p}_{result-interest} + T^{p,i}_{data} + C^i_d + T^{i,c}_{result}
\end{align}
Where $T^{c,p}_{metadata-interest}$, $T^{p,b}_{metadata}$,
$T^{b,p}_{result-interest}$, $T^{p,i}_{data}$, and $T^{i,c}_{result}$ are the
transfer delay of metadata-\Interest, metadata, result-\Interest, \Data, and
result, respectively. Formula \ref{eq:total-delay} indicates that the end-to-end
delay contains two parts. $T^{c,p}_{metadata-interest} + T^{p,b}_{metadata}$ is
the delay of the executor selection of the first phase, and
$T^{b,p}_{result-interest} + T^{p,i}_{data} + C^i_d + T^{i,c}_{result}$ the
result processing delay of the second phase.

To minimize the end-to-end delay $D$, we select the best executor $v_i$. Thus,
our objective function is $\min_{v_i \in V} D$. The cost
$T^{c,p}_{metadata-interest}$ of forwarding the metadata-Interest to the
producer is inevitable. Thus, we further simplify the objective function to:
\begin{align}
  & \min_{v_i \in V} D \Rightarrow \nonumber\\ 
  & \min_{v_i \in V} \{T^{p,b}_{metadata} + T^{b,p}_{result-interest}
  + T^{p,i}_{data} + C^i_d + T^{i,c}_{result}
  \}
\end{align}
\(T^{p,i}_{data}, C^i_d \) and $T^{i,c}_{result}$ are
dominated by the position of executor $v_i$ within the continuum, where $v_i$
is the pivot that contributes to $D$.

We aim to determine $v_i$ to minimize $D$. A traditional method checks all the
nodes in $V$ by estimating the transmission and computation cost in the first
phase. Thus, we write OptimizationOff (OptOff for short), OptOff is a base
version of R2 when $b = c$, i.e., OptOff is a fully two round-trip method.
During the first phase (executor selection), it scans all forwarding nodes.

\noindent\textbf{Cost estimation model.} %
To determine $v_i$, we use a simple cost estimation model that utilizes instant
bandwidth of the forwarding path and CPU cycles of the executor. Let a function
$f$ have a time complexity $\mathcal{O}(\cdot)$ and involve a metadata packet
$M$. In R2, the critical indicators in $M$ are typically \texttt{{datasize,
metasize}} of the data and the metadata packet, respectively. Let $Te_d^{p,i}$
denote the estimated transmitting data delay from $p$ to $v_i$, and $Ce_d^i$
the estimated delay of executing $f$ on $v_i$. Then,
\begin{equation}
  Te_d^{p,i} = M_{datasize} * \frac{M_{metasize}}{(T_{start} - T_{current})}
\end{equation}
and
\begin{equation}
  Ce_d^i = \frac{c\cdot \mathcal{O}(M_{datasize})}{CPU_{frequency}}
\end{equation}
where $M_{datasize}$, $M_{metasize}$, $c$, and $CPU_{frequency}$ are the size of
the original dataset, size of $M$, CPU cycles per operation, and CPU cycles per
second, respectively.

Note that we estimate $Te_d^{p,i}$ in real-time, and thus, other methods can
also be used, e.g., based on historical network interfaces dataset.

Algorithm \ref{alg:opt-off-find-vi} depicts the executor selection steps in the
first phase and from a single node perspective. The algorithm distributively finds
an executor and uses a \texttt{MinCostMarker} stored in the metadata to identify
the executor $v_i$. When the metadata carried a \texttt{MinCostMarker} reaches a
client, the Bolt app running on the client extracts the  \texttt{MinCostMarker}
and attaches it to the result-Interest. This marker is the key to finding the
executor in the second phase, and we preserve this marker until the function is
executed on $v_i$.

\begin{algorithm}
  \caption{OptOff--Executor selection (finding $v_i$)}
  \label{alg:opt-off-find-vi}
  \begin{algorithmic}[1]
    \If {metadata \Data packet}
      \State $etaCost  \leftarrow $  costEstimation($metadata$);\Comment{$ Te_d^{p,i} + Ce_d^i $}
      \State $minCost  \leftarrow $ getMinCost($metadata$);
      \If {$ etaCost \leq minCost $}
        \State $minCost \leftarrow etaCost$;
        \State $MinCostMarker \leftarrow$  hash($metadata$, $node.uuid$);
        \State updateMinCost($metadata$, $minCost$, $MinCostMarker$);
      \EndIf
    \EndIf
  \end{algorithmic}
  \end{algorithm}

However, we argue that \textbf{finding the executor $v_i$ in the second half of
the first phase does not require traveling all nodes in the path, i.e., $v_i
\in$ \{a subset of $V$\}.} In other words, $b$ can be the node along the path
from $c$ to $p$. Next, we propose an optimization called the \textit{stop
condition}.

\subsubsection{Stop condition}\label{sec:stop-condition} %
In the secretary problem \cite{Bearden_2005}, \textit{``a manager sequentially
observes applicants randomly for a single position. When she observes the $b$-th
applicant in the sequence, she learns only the quality of that applicant
concerning those previously seen. Her objective is to select the one who is the
best overall—i.e., relative to all applicants, among those seen and those
not-yet-seen.''}. For the scope of this paper, we redefine this problem as
\textbf{precisely and not probabilistically finding within the continuum the
stop point (the bound $b$) without traveling all nodes in $V$}.

\begin{figure}[H]
  \centering
  \includegraphics[width=.4\textwidth]{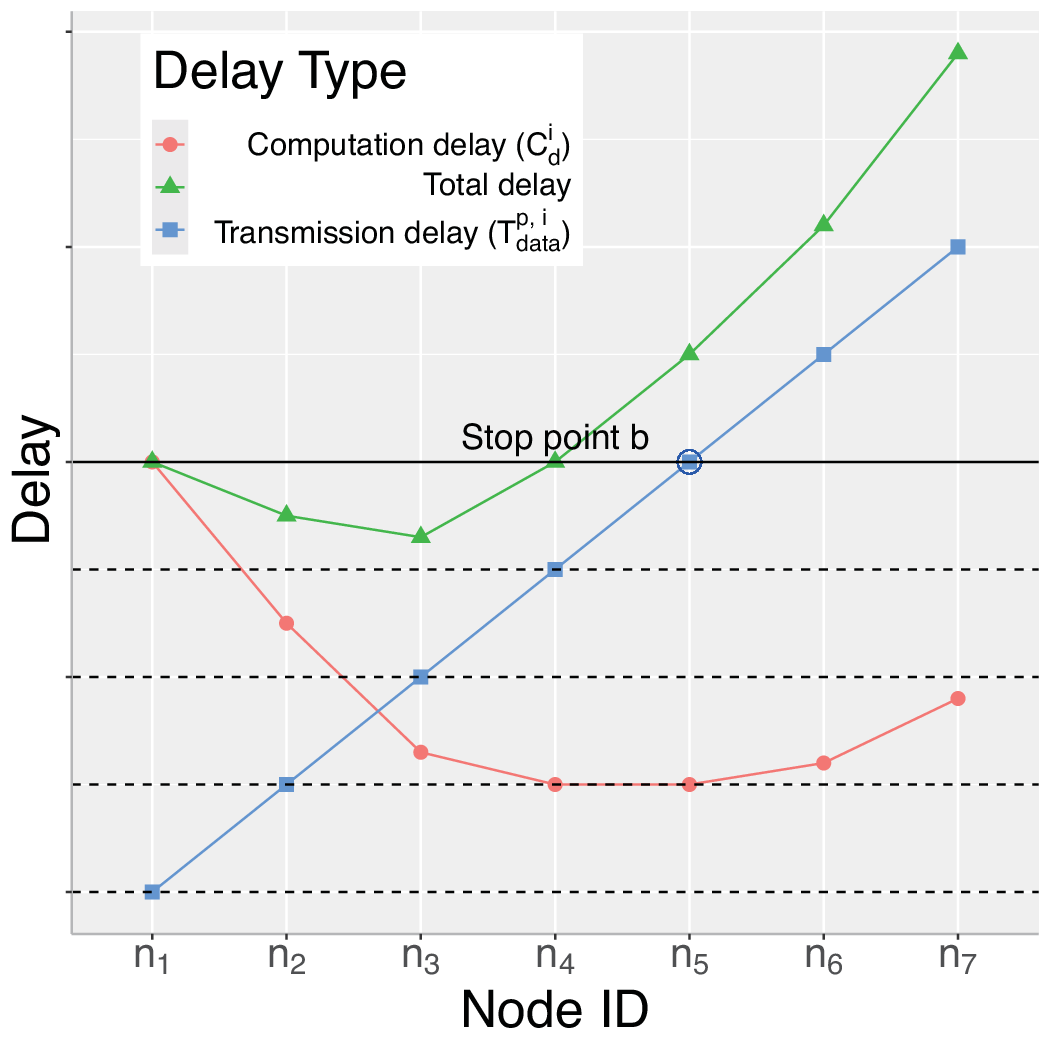}
  \caption{An example of finding the stop point $b$.}
\label{fig:stop-point}
\end{figure}

Based on our findings, cost modeling delays ($T^{p,c/b}_{metadata} +
T^{c/b,p}_{result-interest}$) related to finding the best executor in the OptOff
method can be further reduced by replacing the ``Alice'' node $c$ with a strict
boundary intermediate node $b$. \Fig \ref{fig:stop-point} intuitively depicts an
example of finding bound $b$ that initiates the result-Interest. This figure
presents seven nodes where a computational task with its data is transferred
from $n_1$ to $n_7$, and $n_4$ and $n_5$ are the most potent backbone nodes or
servers. The solid red line shows the computation delay, which is negatively
correlated to the computing power of each node. The solid blue line shows the
\Data transmission delay, and the solid green line shows the total delay. When
the transmission delay $T^{p,i}_{data}$ is greater than the total delay on every
passed-by node, $v_i$ (in this example, $n_3$) is already involved. If we
continuously check the remaining nodes after $n_5$, i.e., $n_6$ and $n_7$, we
obtain only the dominated transmission delay. This phenomenon applies to two
terminals across the backbone network, such as the cross-edge analytic area
\cite{Jin_2021}. Thus, by replacing the marks in the estimation process, formula
\ref{eq:stop-condition} gives the \textit{stop condition}:
\begin{equation}\label{eq:stop-condition}
     Te_d^{p,b} \geq \max_{ 1 \leq j < b}{(Te_d^{p,j} + Ce_d^j)}
\end{equation}
\begin{proof} The stop condition refers to the minimal cost node $n_m$ is
between $n_1$ and $n_b$, i.e., $1 \leq m \leq b$. The proof’s core concept is to
consider $m'$, with $m' \geq b$, to check if its total cost is smaller than node $m$.
Note that the relation of ``$\geq$'' in the stop condition does not change.

Suppose, contrary to our claim, that the stop condition is false. Then we could
find an $m' \geq b$ and $Te_d^{p,m'} \geq Te_d^{p,b}$, imposing a minimal cost
of $Te_d^{p,m'} + Ce_d^{m'}$. Then,
\[
Te_d^{p,m'} + Ce_d^{m'} \leq \max_{ 1 \leq j < b}{(Te_d^{p,j} + Ce_d^j)}
\]
Because  $Te_d^{p,m'} \geq Te_d^{p,b}$, thus
\[Te_d^{p,m'} + Ce_d^{m'} > Te_d^{p,b} \]
Finally, we get
\[
  \max_{ 1 \leq j < b}{(Te_d^{p,j} + Ce_d^j)} < Te_d^{p,m'} + Ce_d^{m'} \leq \max_{ 1 \leq j < b}{(Te_d^{p,j} + Ce_d^j)}
\]
Hence,  \(m' = j < b\),  contradicting our assumption that $m' > b$.When this
condition is met, we stop the executor selection process.
\end{proof}

Based on the stop condition, Algorithm \ref{alg:opt-on-find-vi} presents an
optimized version of Algorithm \ref{alg:opt-off-find-vi}, termed
OptimizationAuto (OptAuto for short).

\begin{algorithm}
\caption{OptAuto--Executor selection (finding $v_i$) within bound $b$}
\label{alg:opt-on-find-vi}
\begin{algorithmic}[1]
  \State {procedures of Algorithm \ref{alg:opt-off-find-vi}} ;
  \If {metadata \Data packet}
    \State $maxCost \leftarrow $ getMaxCost($metadata$);
    \If { $Te_d^{p,i} \geq  maxCost$} \Comment{the stop condition}
      \State TURN INTO second phase; \Comment{$b$ is found}
    \ElsIf {$maxCost < costEta$}
      \State updateMaxCost($metadata$, $maxCost$);
    \EndIf
  \EndIf
\end{algorithmic}
\end{algorithm}

\subsubsection{Applied range analysis of R2}\label{sec:applied-range-analysis}
\textbf{Analyzing the stop condition.} The stop condition can operate
automatically as a plugin. This subsection gives a simple applied range
analysis, especially for the edge where the network and computing are
resource-constrained. It should be noted that $n_j$ is the slowest node
executing the function within the continuum. Thus, we leave out the $max$
symbol, i.e., \(Te_d^{p,b} \geq Te_d^{p,j} + Ce_d^j \). 

The analysis comprises two scenes, a store and forward network (e.g., NDN) and a
universal network (e.g., NDN, TCP/IP).

(1) Store and forward network:
\begin{align*}
&Te_d^{p,b} \geq Te_d^{p,j} + Ce_d^j \implies Te_d^{p,b} - Te_d^{p,j} \geq Ce_d^j \\
&\overset{sn}{ \implies } Te_d^{j,b} \geq Ce_d^j
\end{align*}
where $sn$ represents for a store and forward network. $Te_d^{j,b} \geq Ce_d^j$
indicates that the stop condition operates with the scene where the
computational throughput, i.e., \(datasize/{Ce_d^j}\), of the slowest node $n_j$
exceeds the transmission throughput from $n_j$ to $n_b$. Here, $b$ is the
terminal (e.g., the client).

(2) A universal network:
\begin{align*}
   Te_d^{p,b} \geq Te_d^{p,b} - Te_d^{p,j} \implies Te_d^{p,b}  \geq Ce_d^j
\end{align*}

We eliminate the transmission delay from $p$ to $n_j$ (i.e., $Te_d^{p,j}$) in
the universal network, which is a relatively tight constraint. $Te_d^{p,b}  \geq
Ce_d^j$ indicates that the stop condition operates with the scene where the
computational throughput of the slowest node $n_j$ exceeds the transmission
throughput from $p$ to $n_b$ (or the terminal). This is useful when the delay in
transmitting data from $n_j$ to $n_b$ is challenging to estimate.

\noindent\textbf{R2 vs. 1-round or 1-phase methods. }%
It should be emphasized that R2 has the benefits presented at the beginning of
Section \ref{sec:R2-process}, even if the total delay is bigger than the 1-round
methods in some cases. Nevertheless, if the raw data is limited or the function
is simple, R2 may not outperform the 1-round methods (the methods regarding the
Client or Producer as the executor). Hence, we compare R2 and Client in NDN to
examine the applicability of R2 in terms of the total delay. Without a loss of
generality by assuming $b=c$, R2 involves fully 2-phase operations in the worst
case.

\begin{align*}
 &T_{interest}^{c,p} + T_{data}^{p,c} + C_d^c > D \\
 \implies& T_{data}^{p,c} + C_d^c > T^{p,c}_{metadata} + T^{c,p}_{result-interest}+ \\
 &T^{p,i}_{data} + C^i_d + T^{i,c}_{result} \\
\end{align*}
Because the metadata contains limited information, the metadata size can be
roughly equal to the result-Interest. Then, by
\[
  T^{p,c}_{metadata} + T^{c,p}_{result-interest} \approx 2 T^{c,p}_{interest}
\]
We obtain
\begin{align*}
  \implies& T_{data}^{p,c} + C_d^c > 2 T^{c,p}_{interest} + T^{p,i}_{data} + C^i_d + T^{i,c}_{result}\\
  \implies& T_{data}^{p,c} - T^{p,i}_{data} -  T^{i,c}_{result}  > %
                                         2 T^{c,p}_{interest} + C_d^i - C_d^c \\
  \overset{sn}{ \implies }& T_{data}^{i,c} - T^{i,c}_{result} > 2 T^{c,p}_{interest} + C_d^i - C_d^c \\
\end{align*}
It is clear that the computing power of $n_i$ is bigger than $c$, and thus
$C_d^i < C_d^c$. Writing $\delta = C_d^c - C_d^i$, then
\begin{align*}
\implies T_{data}^{i,c} - T^{i,c}_{result} > 2 T^{c,p}_{interest} - \delta
\end{align*}
Let $\alpha$ be the ratio of the output $result$ size  to the input $data$ size.
\begin{align*}
  \implies (1-\alpha)T_{data}^{i,c} > 2 T^{c,p}_{interest} - \delta
\end{align*}

In most data-intensive functions or applications, $\alpha \ll
1$. Then,
\begin{align*}
  \overset{\sim}{ \implies } T_{data}^{i,c} > 2 T^{c,p}_{interest} - \delta
\end{align*}
From the latter equation, we observe that if $2 T^{c,p}_{interest} < \delta$,
i.e., the client’s computing power is smaller than node $i$ within the
continuum, and the data size is greater than the result. This applied condition
is always true. Second, if the computing resources are similar within the
continuum, then $\delta \approx 0$. For the data size analysis, we obtain
\begin{align}\label{eq:ineq}
  &T_{data}^{i,c} - T^{i,c}_{result} > 2 T^{c,p}_{interest} - \delta \overset{\sim}{ \implies } \nonumber\\
  & T_{data}^{i,c} > 2 T^{c,p}_{interest} + T^{i,c}_{result}
\end{align}
Inequality \ref{eq:ineq} means the data size should be at least greater than the
3-folds of the interest when $i$ is in the vicinity of $p$, which is trivial.

\begin{sidewaysfigure*}
  \vspace*{-1in}
  \centering
  \includegraphics[keepaspectratio,width=.95\textwidth,height=.95\textheight]{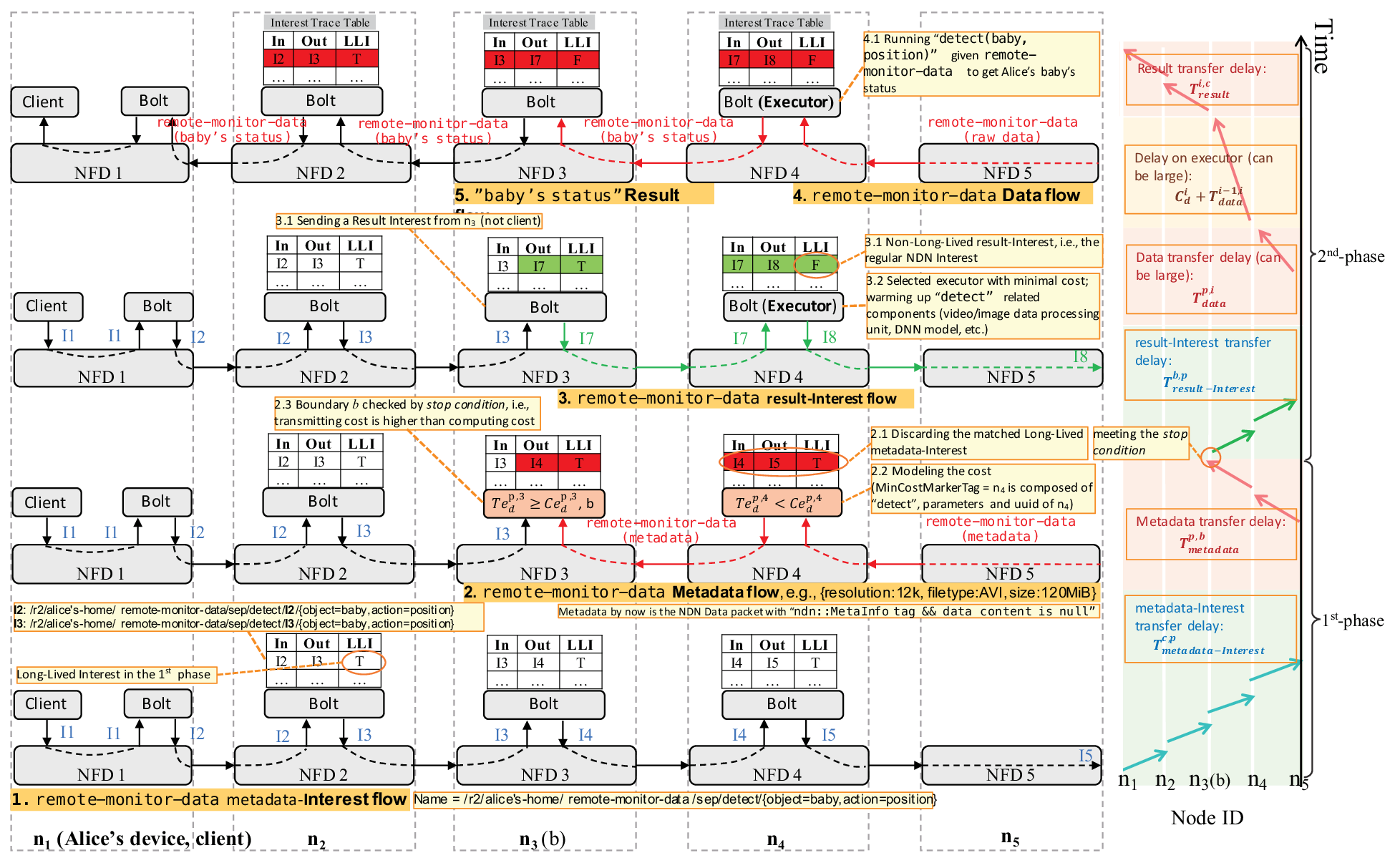}
  \caption{\textbf{R2 process in a step-by-step approach: an auto-mode to select executor.}
  }
  \label{fig:automatically-retrieve-result}
\end{sidewaysfigure*}

\begin{figure*}[h]
  \centering
  \includegraphics[width=.85\textwidth]{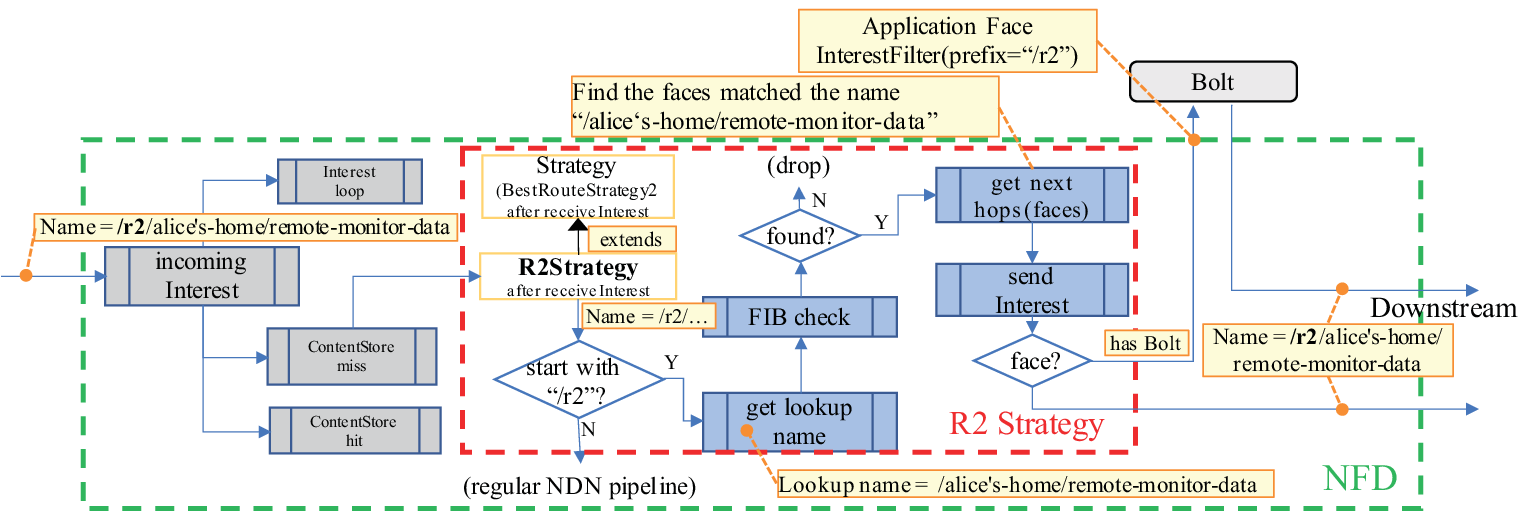}
  \caption{R2 forwarding strategy}
  \label{fig:r2strategy}
\end{figure*}

\begin{figure*}[btp]
  \centering
  \includegraphics[width=.85\textwidth]{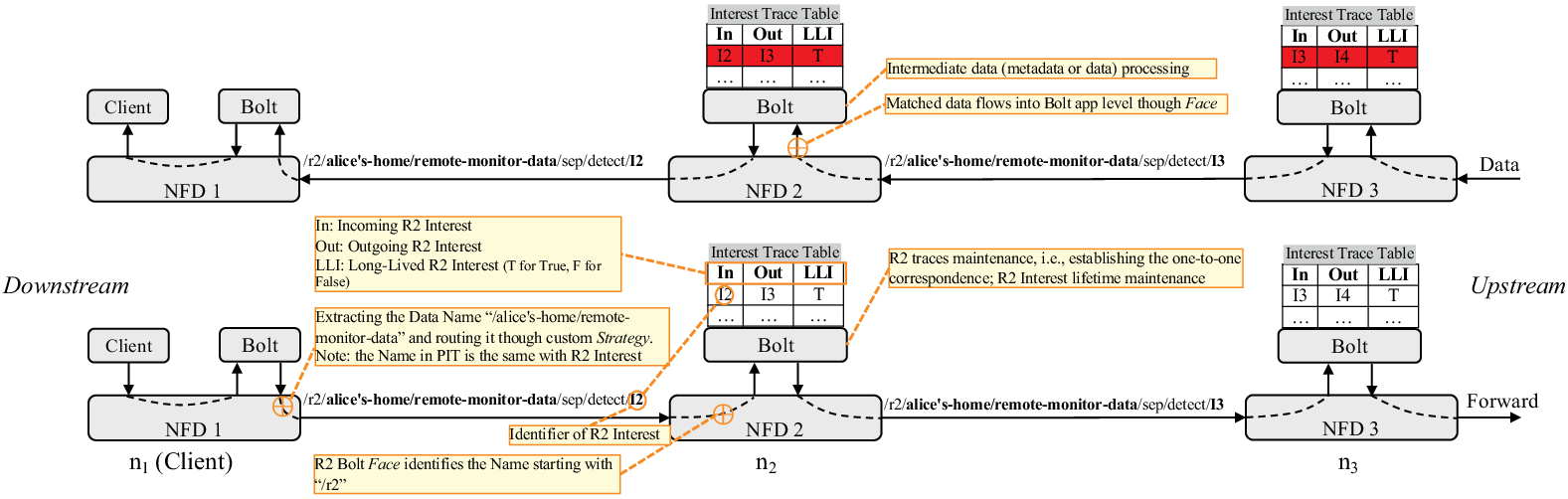}
  \caption{Forwarding Interest/Data using Bolt app.}
  \label{fig:bolt-logic}
\end{figure*}

\subsubsection{R2 process in a step-by-step approach}\label{sec:step-by-step-approach}%
\Fig \ref{fig:automatically-retrieve-result} presents OptAuto by depicting two
phases (comprising five processes or steps) of the R2 protocol under Alice’s
example, who wants to check her baby’s status. By neglecting the stop condition,
R2 becomes a fully 2-phase process. Due to the page size, we involve four
computing-capable nodes ($n_1, \dots, n_4$) and one forwarding node ($n_5$). The
left area in \Fig \ref{fig:automatically-retrieve-result} provides the R2 steps
from the node perspective, while the right area gives a bird’s eye view of the
time delays per step. In Alice's example, she first sends a metadata Interest to
get the remote-monitor-data abstract. R2 automatically creates the traceable
traces, finds the best executor of ``detect'', retrieves raw data of
remote-monitor-data, analyzes her baby’s status, and finally sends the baby’s
status to Alice. We describe these five steps in detail below.

(1) Alice first sends a long-lived metadata-\Interest (under the name
\texttt{/r2/alice's-home/remote-monitor-data/sep/
detect/\{object=baby,action=position\}}) into NDN.  This Interest passes through
the nodes within the continuum, is captured by Bolt, leaves a mapping
correspondence in ITT, and finally reaches the producer (or the camera at
Alice’s home). The time consumed $T_{metadata-Interest}^{c,p}$ in this process
is positively correlated with the packet size of the metadata-Interest and the
number of hops between Alice and her home.

(2) The producer first replies with a metadata \Data packet containing
\texttt{\{resolution:12k, filetype:AVI, size:120MiB\}}. When the metadata moves
towards Alice, two critical actions are performed. The first is to find the
executor with a minimal end-to-end delay cost $D$ based on node configuration,
``detect’’, and metadata. According to the stop condition, the second is to find
the boundary $b$, i.e., $n_3$. R2 discards the created mapping correspondence
(red records) to reduce the ITT size in this process. After this process, we
obtain a \texttt{MinCostMarker} representing the executor, i.e., $n_4$. This
marker comprises ``detect'', parameters and the node’s UUID, which can be viewed
as the identifier of Alice’s request. The time consumed $T_{metadata}^{p,b}$ in
this process is positively correlated with the packet size of metadata and the
cost estimation step, which is insignificant.

(3) $b$ issues a result-Interest carrying the \texttt{MinCostMarker} to the
producer. Since all we need is the raw data, i.e., the HD video file
remote-monitor-data, any other operations during transmitting the raw data
between $b$ and the producer are needless. Thus, the result-Interest of the
newly added mapping correspondence in ITT (green records) is set to be the
Non-Long-Lived Interest type. We can also perform warm-up services, e.g.,
video/image data processing unit, DNN model, and libraries related to performing
\texttt{detect(baby, position)}, on $n_4$ when it receives the notification from
$b$ to avoid a cold start. The time consumed $T_{result-Interest}^{b,p}$ in
this process is positively correlated with the packet size of result-Interest,
the number of hops between the boundary and Alice’s home. For completeness, it
should be mentioned that the path guided by the NDN routing table (or PIT),
despite being unstable, it can be settled through session support
\cite{GasparyanMarandiSchillerEtAl2019}.

For the step (4) and (5), the executor (i.e., $n_4$) runs the function after it
receives the Data to provide Alice’s baby’s status. Finally, the result is
returned to the client. The time consumed in processes 4 and 5 comprise and are
positively correlated with three parts: result transfer delay
$T_{result}^{i,c}$, computation delay $C_d^i$ and raw data transfer delay
$T_{data}^{p,i}$.

\section{Evaluation and results}\label{sec:evaluation-and-results}%
To evaluate R2's performance, we design an intermediate data processing logic
on the ``Bolt'' application level and implement it in ndnSIM. This section
first illustrates the ``Bolt'' logic (Section \ref{sec:r2-bolt}) and then
exploits it to evaluate R2 (Section \ref{sec:evaluation}).

\subsection{R2 Bolt: intermediate data processing on app level}
\label{sec:r2-bolt}
Automatically running a function requires a modular implementation not only for
maintenance but also for development. A general forwarding framework named
NDN-trace \cite{KhoussiPesaventoBenmohamedEtAl2017} processes the intermediate
Interest and Data. From a scalability and compatibility perspective, we extend
NDN-trace and perform a custom forwarding strategy to support function execution
on an application level, i.e., Bolt\footnote{The name ``Bolt'' is inspired by
Apache Storm. In Apache Storm, the logic for processing data tuples in a node is
called a ``Bolt''.}. Next, we provide the R2 Bolt app details.

\Fig \ref{fig:r2strategy} depicts the custom forwarding strategy pipeline of R2.
The red and green dotted lines frame the R2 Strategy and the NFD (the NDN
Forwarding Daemon). When NFD receives an R2 Interest, it is forwarded according
to the FIB (Forwarding Information Base) to the Bolt of that node. Inside NFD,
R2 Interests are handed over to our custom R2 Strategy extended from the native
\texttt{BestRouteStrategy2} of NDN. R2 Strategy first performs a next-hop lookup
on the actual data name \texttt{/alice’s-home/remote-monitor-data} only,
extracted by the ``\texttt{get lookup name}'' method instead of the entire R2
Interest name with the prefix ``/r2''. Then, after checking the FIB, we directly
obtain the next hop (or face) matched to the name
\texttt{/alice’s-home/remote-monitor-data} by performing the method ``get next
hops''. At last, \texttt{/r2/alice’s-home/remote-monitor-data} is directly sent
to the matched next hop (or face). Note that the next hop (or face) contains two
types: Bolt and the next hop node. Because the Bolt app registers the prefix
``\texttt{/r2}'' using \texttt{InterestFilter}, it can directly identify and
capture the R2 Interest. If the node does not install the Bolt or is a
computing-incapable node R2 Strategy, it will forward
\texttt{/r2/alice’s-home/remote-monitor-data} to the direction of
\texttt{/alice’s-home/remote-monitor-data} directly. The entire process is
repeated at every node encountered on the path(s) until R2 Interest
\texttt{/alice’s-home/remote-monitor-data} reaches the target prefix, i.e.,
Alice's home.

\Fig \ref{fig:bolt-logic} depicts a forwarding pipeline using three
computing-capable nodes to demonstrate the Bolt logic. To simplify the process,
we do not distinguish between Metadata and Data. Alice ($n_1$) sends an Interest (under the
name \texttt{/r2/alice's-home/remote-monitor-data/sep /detect/paras}) into the
upstream. Its true data name \texttt{/alice's-home/remote-monitor-data} is first
extracted by a custom \texttt{Strategy} implemented at the NFD level, and then
it is forwarded to the next hops according to the forwarding policy by scanning
FIB\footnote{It is worth noting that R2 Interest starts with a prefix
\texttt{/r2} (\Fig \ref{fig:interest-packet-structure}), while the forwarding
process of NFD
only cares about \texttt{/alice's-home/remote-monitor-data}. Thus, the right
sub-components in the names are extracted through custom \texttt{Strategy} in the NFD
level to enter the proper forwarding pipeline. The Interest Name in PIT is still
starts with \texttt{/r2} and does not change.}. %
When the incoming R2 Interest is received by the NFD, which has Bolt installed,
it is caught by Bolt  \texttt{Face} and transmitted into the processing logic
(\texttt{processingInterest} in \Fig \ref{fig:process-incoming-interest}).
Currently, Bolt has an Interest Trace Table (ITT) performing three actions: (1)
Clones the incoming Interest and append a random number (Interest identifier,
e.g., I1, I2,….) to the tail of the cloned Interest \texttt{Names}. (2) Creates
a mapping entry between the cloned and the origin Interest and inserts the entry
into ITT. (3) Deletes the entry if the Interest is satisfied (red records in
ITT). Additionally, since the function execution process may take time, another
important role of ITT is manipulating the lifetime (Long-Lived Interest flag in
ITT) of the R2 Interest to avoid the time-out issue imposed by solely employing
PIT. As the ndnSIM declares, these codes can be applied to the production with
minor modifications.

\begin{figure}[!hbtp]
\begin{lstlisting}[language=C++]
void onInterest(const InterestFilter& filter, const Interest& inInterest) 
{
  // Discard looped Interest
  if (Islooped(inInterest)) return;
  // Clone the incoming Interest and create the mapping relationship
  preProcessingInterest(inInterest, outInterest);
  // Process the cloned Interest (i.e., outInterest)
  processingInterest(filter, inInterest, outInterest);
  // Issues the cloned Interest
  afterProcessingInterest(inInterest, outInterest);
}
\end{lstlisting}
\caption{Pseudo-code of processing an incoming Interest in Bolt: a single node perspective}
\label{fig:process-incoming-interest}
\end{figure}

The pseudo-code to process an incoming \Interest is depicted in \Fig
\ref{fig:process-incoming-interest}. The parameter \texttt{inInterest} can be
referred to as \texttt{I1, I2}, and \texttt{I3} (start with a prefix \texttt{/r2})
that flows into Bolt in \Fig \ref{fig:bolt-logic}. Line 6 first clones those
incoming \texttt{Interest}s then inserts the mapping relationship into ITT, and
finally exits  the cloned \texttt{outInterest}. Sending out the \texttt{outInterest}
(Line 10) refers to \texttt{I2, I3} and \texttt{I4} that exit Bolt in \Fig
\ref{fig:bolt-logic}. In the \texttt{processingInterest} (Line 8), some additional
operations can be applied to \texttt{outInterest}, e.g., make the interest
long-lived, or add user-interested tags. In the R2 Bolt, processing the \Data is
technically the same as processing \Interest logic. For further details, the
reader is referred to the source code\footnote{Intermediate data processing code
in Bolt: https://git.io/J3iux}.

\subsection{Evaluation}
\label{sec:evaluation}
\begin{figure}[hbtp]
  \centering
  \includegraphics[width=75mm]{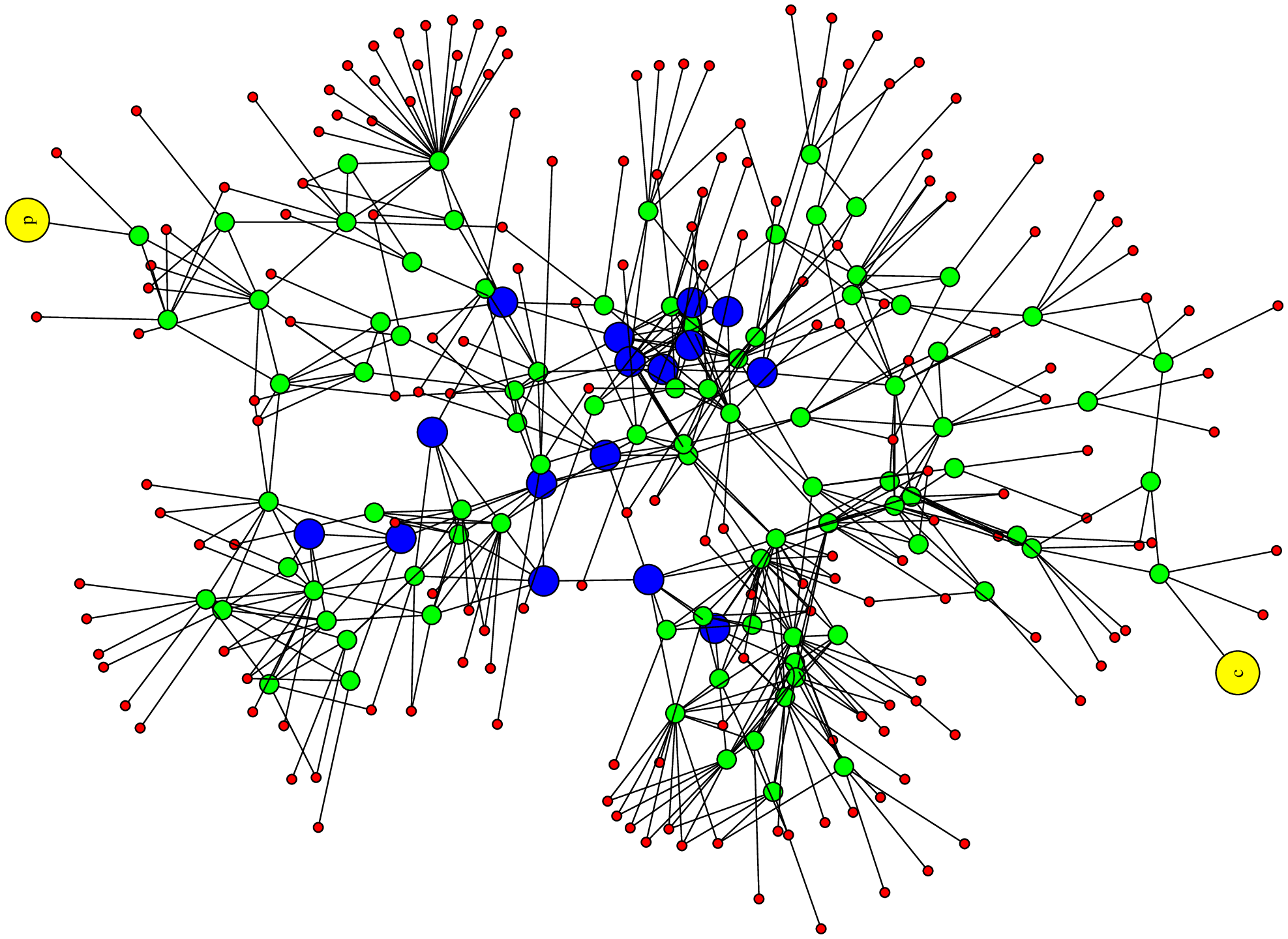}
  \caption{Rocketfuel dataset.}
\label{fig:rocketfuel-topology}
\end{figure}

This section evaluates R2 via an extensive simulation study. We implement four
notable methods, including OptOff and OptAuto, to testify the end-to-end delay
in ndnSIM \cite{mastorakis2017ndnsim}, where ndnSIM is an NS-3-based simulator
for NDN \cite{ZhangAfanasyevBurkeEtAl2014} implementation. Two additional
methods are the ``Producer'' and ``Local'', where the former means pushing the
function down to the storage (data location), a typically distributed database
management method. However, the Producer's computing power and energy are
usually limited. The Local method considers the client directly retrieving the
\Data and executing the function itself. All R2 codes are based on ndnSIM-2.8,
NFD-0.7.0, and all the experiments are performed on macOS utilizing an Intel i5
Dual-core at 2.7GHz. 

\textbf{Dataset.} We adopt a real-world network topology, i.e., rocketfuel
\cite{SpringMahajanWetherallEtAl2004}, comprising 282 nodes, including 177
clients, 89 gateways, and 16 backbones. The topology is illustrated in \Fig
\ref{fig:rocketfuel-topology}, where the two yellow nodes are the client (c) and
the producer (p). The blue backbone is the high-level performance edge node, the
green gateway is the middle-level performance edge node, and the red device is
the low-level edge node performance. Table \ref{tab:nodes-configuration} shows
this topology's configurations, where H and M denote the high and middle
performance, respectively, and C the end device (Client or Producer). It is
worth noting that the available bandwidth is the total bandwidth split by
numerous nodes due to concurrency and collision.

\begin{table}[!th]
\renewcommand{\arraystretch}{1.3}
  \begin{adjustwidth}{0in}{0in} 
  \centering
  \caption{Topology configuration}
  \label{tab:nodes-configuration}
  \begin{center}
    \begin{tabular}{c c c c} 
      \hline
      Item & H(-H) & H(-M) & M(-C) \\ [0.5ex] 
      \hline\hline
      Available bandwidth (Mbps) & 10 $\sim$ 40 & 20 $\sim$ 40 & 40 $\sim$ 100 \\ 
      \hline
      CPU (GHz) & 3  $\sim$ 4  & 2  $\sim$ 3  & 0.5  $\sim$ 2  \\
      \hline
      Memory (GB)& 32 $\sim$ 128 & 8 $\sim$ 32 & 2 $\sim$ 8 \\
      \hline
   \end{tabular}
   \end{center}
  \end{adjustwidth}
\end{table}

\textbf{Indicators.} We focus on measuring the average end-to-end delay and the
hop-by-hop delay. End-to-end delay is used to compare four methods, and the
hop-by-hop delay is used to test whether the stop condition is working.
Different complexity of the function can have a different impact on the cost
estimation model. Thus, given the competitor methods, we categorize the time
complexity of the function as $\mathcal{O}(logn)$, $\mathcal{O}(n)$, and
$\mathcal{O}(n^2)$ and fixed space complexity as $\mathcal{O}(n)$ \footnote{For
further information on IOPS the reader is referred to:
https://docs.aws.amazon.com/AWSEC2/latest/UserGuide/ebs-io-characteristics.html},
where $n$ is the data size. In the simulation, we set the ratio of the function
output size to its input data size as $\alpha=0.1$.

\subsection{Results}

\begin{figure}[!th]
  \centering
  \includegraphics[width=.4\textwidth]{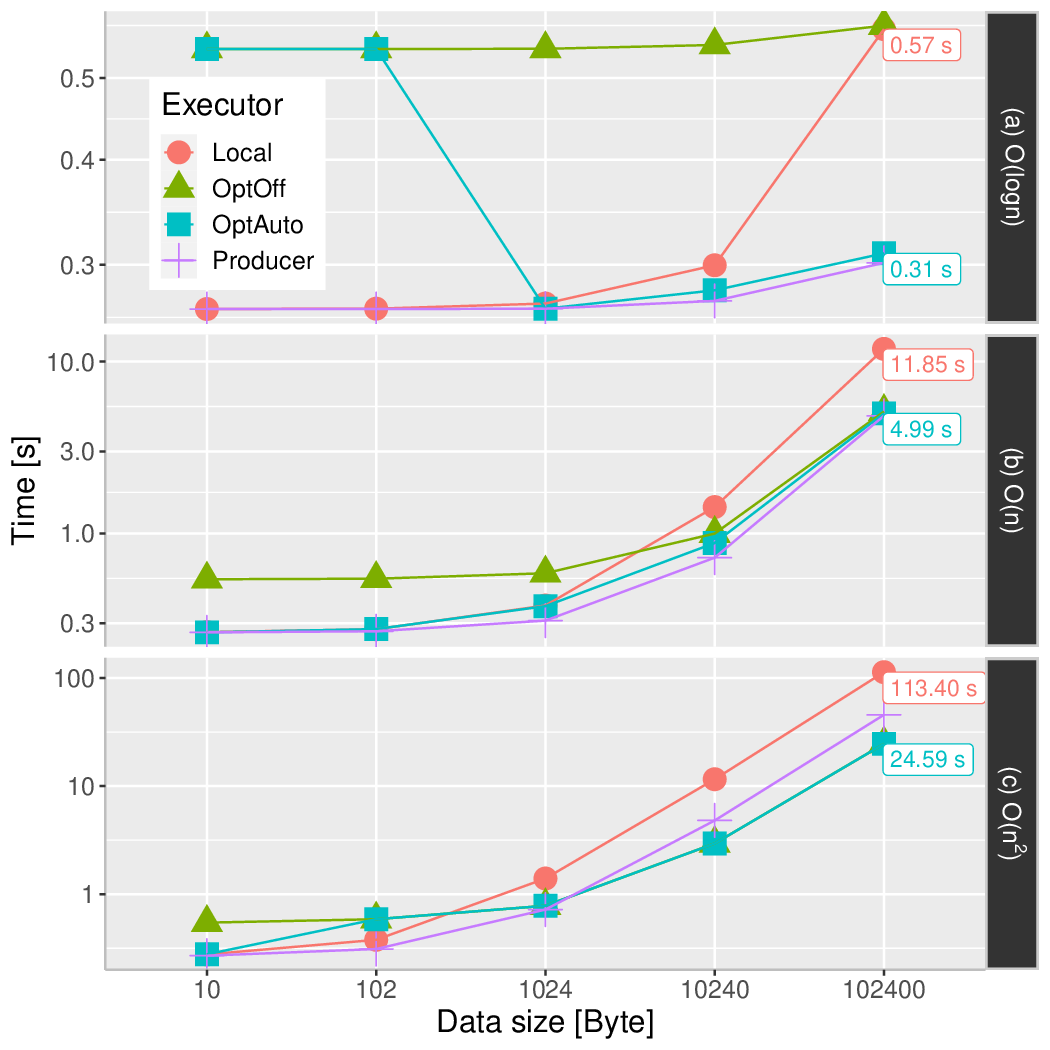}
  \caption{End-to-end delay of 4 methods. 
  }
  \label{fig:end-to-end-daly-total}
\end{figure}

\Fig \ref{fig:end-to-end-daly-total} presents the end-to-end delay of the four
evaluated methods by varying the data size and time complexity. Fig. 9 indicates
that the end-to-end delay increases as the data size and complexity rise. When
the data size is small, i.e., data size $<$ 102 bytes, both the Local and
Producer present a low delay level. However, when the data size exceeds 1024
bytes (1 KiB, which is very rare in today’s network communication), OptAuto and
OptOff gradually get better than the Local and Producer. Hence, the executor is
essential for the computations within the continuum.

\Fig \ref{fig:end-to-end-daly-total} (a) illustrates that  OptOff, which has a
complexity of $\mathcal{O}(logn)$ , is always at a high level because it needs a
fully two-round-trip to find the executor and the transmission dominating the
total delay. Since the $\mathcal{O}(logn)$ operations are not large, the
Producer is faster than the competitor methods. Nevertheless, we do not prefer
the task compute at the producer because its computing power is usually limited.
Meanwhile, OptAuto obtains a similar result with the Producer when the data size
is larger than 1024 bytes, i.e., the bound $b$  is found, and a fully
two-round-trip compared with OptOff is avoided. Local also presents an increased
processing time once the data size exceeds 1024 bytes, which integrating with
OptAuto implies that the computation and the transmission delays are of the same
order and both dominate the total delay. It should be noted that the methods
having $\mathcal{O}(logn)$ complexity are not everywhere reachable, especially
in the data analysis realm. \Fig \ref{fig:end-to-end-daly-total} (b) and \Fig
\ref{fig:end-to-end-daly-total} (c) the four methods with $\mathcal{O}(n)$ and
$\mathcal{O}(n^2)$, respectively. From this figure, it is evident that OptAuto
manages an acceptable result compared with Local and Producer. When the data
size reaches 100 KiB, the speedup ratio Local/OptAuto reaches 1.84, 2.37, and
4.61 for $\mathcal{O}(logn)$, $\mathcal{O}(n)$, and $\mathcal{O}(n^2)$,
respectively.

\begin{figure}
  \centering
\begin{subfigure}[ht]{0.45\textwidth}
  \includegraphics[width=\textwidth, height=.8\textwidth]{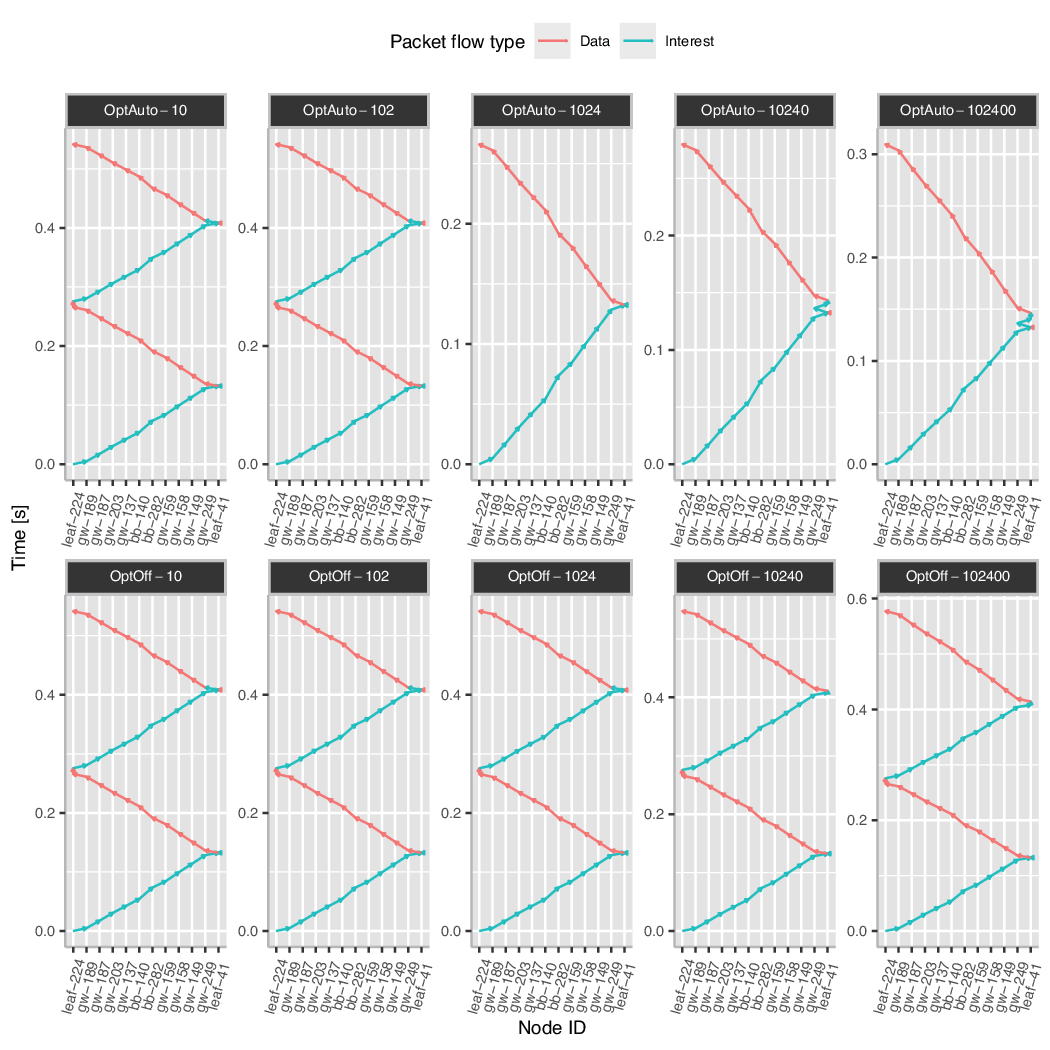}
  \caption{\bf{Hop-by-hop accumulating delay of $\mathcal{O}(log(n))$.}}
  \label{fig:hops-delay-logn}
\end{subfigure}
\begin{subfigure}[ht]{0.45\textwidth}
  \includegraphics[width=\textwidth, height=.8\textwidth]{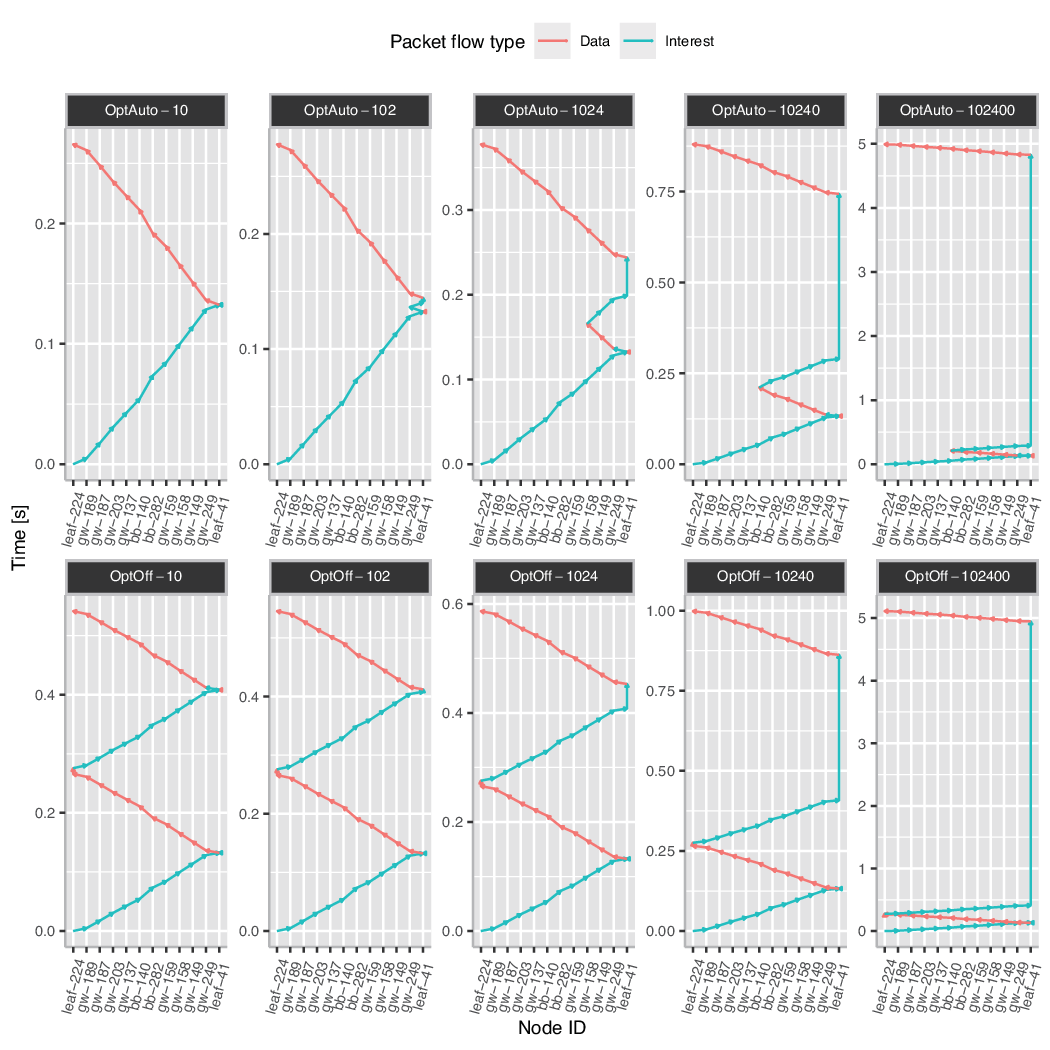}
  \caption{\bf{Hop-by-hop accumulating delay of $\mathcal{O}(n)$.}}
  \label{fig:hops-delay-n}
\end{subfigure}
\begin{subfigure}[ht]{0.45\textwidth}
  \includegraphics[width=\textwidth, height=.8\textwidth]{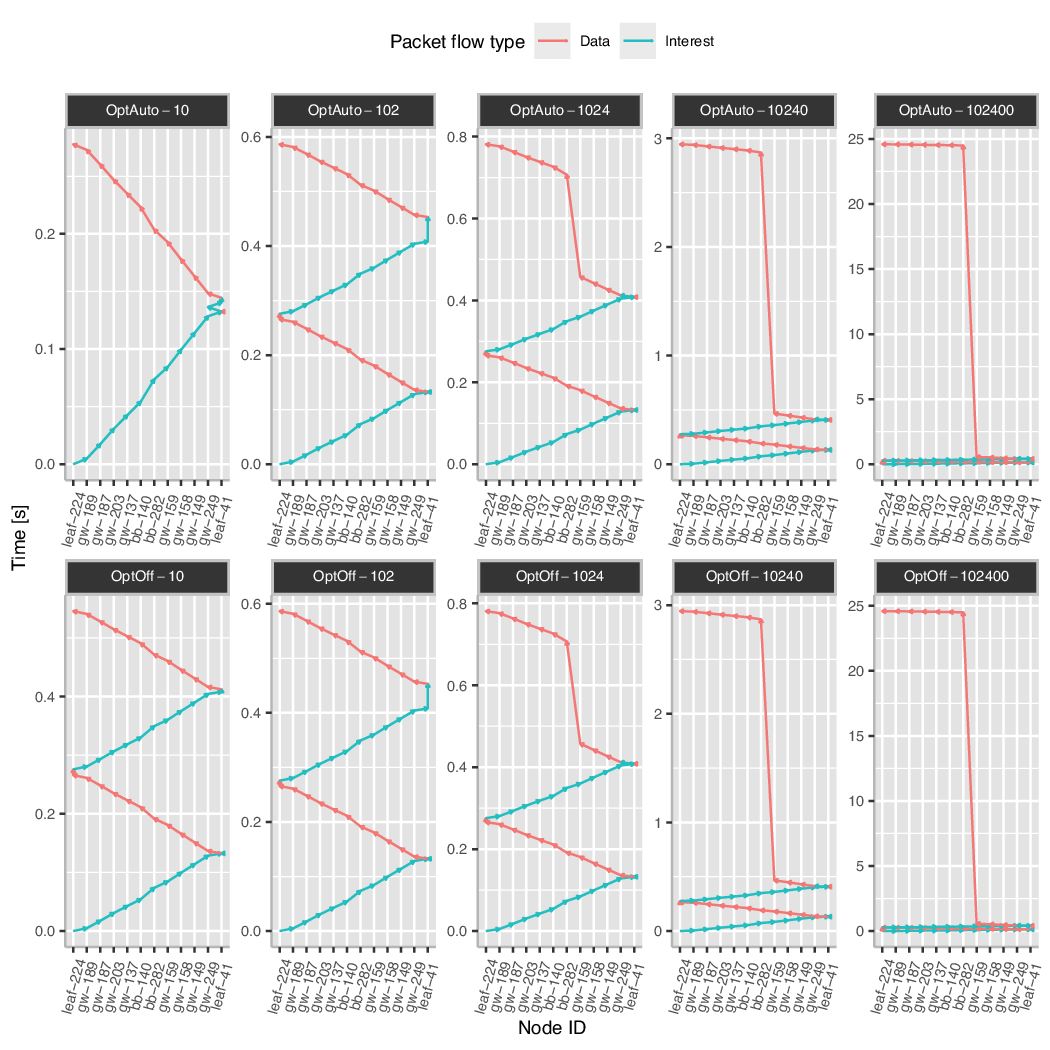}
  \caption{\bf{Hop-by-hop accumulating delay of $\mathcal{O}(n^2)$.}}
  \label{fig:hops-delay-n2}
\end{subfigure}
\caption{Hop-by-hop accumulating delay of OptAuto and OptOff varying data size}
\label{fig:hop-by-hop-delay}
\end{figure}

From the stop condition analysis of Section \ref{sec:stop-condition}, we know
that the bound $b$ is related to the time complexity and transmission delay.
Next, to verify whether the bound $b$ is found, we plot the hop-by-hop
accumulating delay of OptAuto and OptOff varying data size (entitled
``Method-Datasize''), presented in \Fig \ref{fig:hop-by-hop-delay}. It is
straightforward that in most cases, OptAuto finds the bound $b$ compared to the
OptOff in most cases, OptAuto finds the bound $b$ compared to the OptOff. Most
importantly, OptAuto gives the same executor as OptOff, proving the
effectiveness of the stop condition that does not check all nodes.

\section{Discussion}\label{sec:discussion} %
The reader might have security-related concerns for the Interest and Data
intermediate processing. However, this is a well-discussed topic in
\cite{ZhangAfanasyevBurkeEtAl2014} and
\cite{KhoussiPesaventoBenmohamedEtAl2017}. In addition to ICN’s content-based
security model, we implement our solution on an application and not system
level, and thus many possible ways can be utilized to enable authentication
services. Feasible solutions involve nodes within the continuum having the
public key of each other or other decentralized authentication methods
\cite{LiuNingLi2005}. Neverthless, security concerns are out of this paper's
scope, and thus we will not further examine them.

This version of R2 focuses on how to pick the best computing-capable executor
within the forwarding path. Thus, mobility is not paid much attention. Recent
mobility-related works mainly focus on two types of nodes (consumer and
producer) and three categories (mapping-based, tracing-based, and data depot)
\cite{MobilitySurvey2016, FengZhouXu2016, kite2018}. For R2, mobility support
needs to be built on top of these ideas. Specifically, in R2, the executor is
the role of the consumer compared to the data producer, and the role of the data
producer remains unchanged. Thus, if the producer moves, the methods mentioned
in \cite{MobilitySurvey2016, FengZhouXu2016, kite2018} can be adapted. %
Another problem is that intermediate forwarding nodes have mobility. Current
works have not paid much attention to the intermediate forwarding nodes, but it
is more common in self-organizing and stochastic networks. For the consumer, the
executor is the role of the producer that may use a tracing-based approach to
proactively inform the nodes on the forwarding path to establishing a new
channel, which ensures that the executor is always accessible. Another solution
is to carry the unique identifier of the executor as a hint when the executor
performs the process of returning ACK Data to the consumer, which ensures that
we can eventually look up the executor.

\section{Conclusion}\label{sec:conclusion}%
This paper discusses the feasibility of R2, a 2-phase novel mechanism for data
processing in edge computing. A 2-phase design has many potential benefits,
ensuring the function execution integrity, checking the node’s health status in
the client's path towards the producer, and, most importantly, choosing the best
node to execute the function. R2 leverages the NDN paradigm for data retrieval
from data sources, enhances it to provide a decentralized method invocation
mechanism with the objectives to (1) minimize the end-to-end delay by limiting
the raw data traffic crossing the network and selecting the executor and (2)
reduce the first round trip time by an intelligent stop condition. In addition,
we developed ``Bolt'' to process the intermediate data on the app level in ndnSIM.
``Bolt'' can be installed on the selected computing-capable nodes to provide
scalable service. We believe that our freely available code \cite{r2sourcecode}
can help researchers and developers verify their ideas smoothly.

\section{Future works}
Currently, R2 only handled a single user request in this paper. We solve the
multi-user requests, including ``compute reuse'' in the future. Future works also
include extending R2 to handle application (i.e., solving the functions/tasks
directed acyclic graph (DAG)), re-forwarding the user’s Interest according to
function name by the intermediate nodes to scale out the executor selection
space, storing metadata and data with different locations.

Changing metadata-Interest's lifetime from long-lived to regular is another
future work. We think the latter is feasible and is currently implemented by
inflating the timer of the metadata-Interest entry in the PIT through sending a
refreshing Interest or an ACK Data that carries the estimated function-execution
time from the selected executor back to the consumer. This is similar to RICE's
``Interest Acknowledgements'' \cite{KrolHabakOranEtAl2018}. The most significant
difference is that the refreshing action in R2 is sent from the executor and not
from the consumer or the producer. Because the executor has the metadata, R2
affords a more precise function-execution time.

Instead of identifying and capturing the current R2 Interest by the prefix
``/r2'', effectively capturing the Interests at the application level by
identifying the user-selected Name components is another future research
direction. Note that our purpose is still putting the computation logic on the
application level, not the NFD level. Another possible solution is adding
``/r2'' as a postfix, identifying the postfix ``/r2'' and forwarding it to the
face of the Bolt app instead of the next-hop node, which still integrates with
the forwarding Strategy. This Strategy seems similar to the Interests tracing
method KITE \cite{kite2018} and opposes the current R2 Strategy. To the best of
our knowledge, designing a scalable naming convention is a concern of many
recent works that put the function or service identifier at the head of the
Name, e.g., name as a function. Overcoming this issue may bring R2 scalability
into a new stage.

\ifCLASSOPTIONcompsoc
  \section*{Acknowledgments}
\else
  \section*{Acknowledgment}
\fi
This work was supported by National Natural Science Foundation of China under
Grant No. 62173158 and No. 61379134 and the Scientific Technological Innovation
Foundation of Shunde Graduate School, USTB under Grant No. BK19CF010 and No.
BK20BF012, and in part by the National Key Research and Development Program of
China under Grant No. 2016YFC0901303.



\bibliographystyle{IEEEtran}
\bibliography{IEEEabrv,library}
\end{document}